\documentclass[aps,showpacs,preprintnumbers,amsmath,amssymb]{revtex4}

\usepackage{graphicx}
\usepackage{dcolumn}
\usepackage{bm}

\begin{document}

\title{Possible Phases of the Two-Dimensional $t-t'$ Hubbard Model}

\author{V.~Hankevych$^{1,2}$}
\author{F.~Wegner$^1$}%
\affiliation{\mbox{$^1$Institut f\" ur Theoretische Physik,
Universit\" at Heidelberg, Philosophenweg 19, D-69120 Heidelberg,
Germany}\\ $^2$Department of Physics, Ternopil State Technical
University, 56 Rus'ka St., UA-46001 Ternopil, Ukraine}

\date{\today}

\begin{abstract}
We present a stability analysis of the 2D $t-t'$ Hubbard model on a
square lattice for various values of the next-nearest-neighbor hopping
$t'$ and electron concentration. Using
the free energy expression, derived by means of the flow equations
method, we have performed numerical calculation for the various
representations under the point group $C_{4\nu}$ in order to determine
at which temperature symmetry broken phases become more favorable
than the symmetric phase. A surprisingly large number of phases has
been observed. Some of them have an order parameter with many nodes in
${\bf k}$-space. Commonly discussed types of order found by us are
antiferromagnetism, $d_{x^2-y^2}$-wave singlet superconductivity,
$d$-wave Pomeranchuk instability and flux phase. A few instabilities
newly observed are a triplet analog of the flux phase, a particle-hole
instability of $p$-type symmetry in the triplet channel which gives
rise to a phase of magnetic currents, an $s^*$-magnetic phase, a
$g$-wave Pomeranchuk instability and the band splitting phase with
$p$-wave character. Other weaker instabilities are found
also. A comparison with experiments is made.
\end{abstract}

\maketitle

\section{\label{sec1}Introduction}

In recent years the two-dimensional (2D) Hubbard model has been
used~\cite{iz,sc} as the simplest model which maps the electron
correlations in the copper-oxide planes of high-temperature
superconductors since experimental data suggest that superconductivity
in cuprates basically originates from the CuO$_2$
layers~\cite{tk}. Although in the high-temperature cuprate
superconductors electron-electron interactions are strong some
important features of these systems (in particular, antiferromagnetic
and $d$-wave superconducting instabilities) are captured already by
the 2D Hubbard model at weak to moderate Coulomb coupling.

Apart from the antiferromagnetism and $d_{x^2-y^2}$-wave
superconductivity above mentioned (for review see~\cite{sc,ha} and
references therein), a few other instabilities related to
symmetry-broken states~\cite{ko,am,hm,foh,hsg,ikk,hs,hsr,gkw,hgw}, and
truncation of the Fermi surface~\cite{hsfr} in the 2D $t-t'$ Hubbard
model with next-nearest-neighbor hopping $t'$ have been reported. One
of these phases is the staggered flux phase of singlet type with
$d$-wave character which occurs in the particle-hole channel and
breaks translational and time-reversal symmetries.  The singlet flux
phase has been discovered in the mean-field treatments of the 2D
Hubbard model by Kotliar~\cite{ko}, and independently by Affleck and
Marston~\cite{am}. It has been discussed by Nayak~\cite{na} and
Chakravarty et al.~\cite{clmn} as a $d$-density wave state which may
coexist with $d$-wave superconductivity. Recently, a triplet analog of
the staggered flux phase has been observed~\cite{hgw} also in the 2D
$t-t'$ Hubbard model in certain regions of the parameters.

Another instability in the 2D $t-t'$ Hubbard model is a $d$-wave
Pomeranchuk instability breaking the tetragonal symmetry of the Fermi
surface, i.e. a spontaneous deformation of the Fermi surface reducing
its symmetry to orthorhombic. It has been recently observed for small
values of $t'$ from renormalization group calculations by Halboth and
Metzner~\cite{hm}. They argued that the Pomeranchuk instability occurs
more easily if the Fermi surface is close to the saddle points of the
single particle dispersion (Van Hove filling) with a sizable $t'$
(reducing nesting which leads to antiferromagnetism). However, within
their technique it is difficult to compare the strength of the Fermi
surface deformation with other instabilities and to conclude which one
dominates. The authors of Ref.~\cite{hsr} have investigated the
interplay of $d$-density wave and Fermi surface deformation tendencies
with those towards $d$-wave pairing and antiferromagnetism by means of
a similar temperature-flow renormalization group approach. They have
found that the $d$-wave Pomeranchuk instability is never dominant in
the 2D $t-t'$ Hubbard model (even under the conditions mentioned
above). At the same time, the $d_{x^2-y^2}$-wave Pomeranchuk
instability has been observed~\cite{gkw} to be one of the strongest
instability in the 2D Hubbard model by means of the flow equations
method, but these calculations have some limitations at low
temperatures (see Section~\ref{sec2}).

On the other hand, Vollhardt et al.~\cite{vbh} showed that
$t'$-hopping destroys antiferromagnetic nesting instability at weak
interactions in two and three dimensions, and supports the
stabilization of metallic ferromagnetism in infinite dimensions away
from half-filling. Therefore, one may expect also the stabilization of
ferromagnetism by a sizable $t'$ in two dimensions.  Indeed, in the
$t-t'$ Hubbard model on a 2D square lattice at week to moderate
Coulomb coupling, a projection quantum Monte Carlo calculation with
$20\times 20$ sites and the $T$-matrix technique~\cite{hsg}, a
generalized random phase approximation including particle-particle
scattering~\cite{foh} point towards a ferromagnetic ground state for
large negative values of $t'$ in a density range around the Van Hove
filling. Similar tendencies have been found by the authors of
Ref.~\cite{ikk} within the renormalization group and parquet
approaches. Honerkamp and Salmhofer recently studied~\cite{hs} the
stability of this ferromagnetic region at finite temperatures by means
of the temperature-flow renormalization group technique. They have
found that ferromagnetic instability is the leading one at $t'<-0.33t$
and the Van Hove filling with critical temperatures depending on the
value of $t'$. When the electron concentration is increased slightly
above the Van Hove filling, the ferromagnetic tendencies get cut off
at low temperatures and a triplet $p$-wave superconducting phase
dominates. However, they did not consider the Pomeranchuk instability
and other ones apart from antiferromagnetism, $d$- and $p$-wave
superconductivity and ferromagnetism.

Other instabilities recently found in the 2D Hubbard model are an
``insulating spin liquid state''~\cite{hsfr} and ``band splitting''
phase~\cite{gkw}. Insulating spin liquid phase is the state with
truncated Fermi surface and unbroken translational symmetry. It is
insulating due to the vanishing local charge compressibility, and a
spin liquid because of the spin gap arising from the pairing
correlations. Band splitting phase is a particle-hole instability of
singlet type with $p$-type symmetry which has been found~\cite{gkw}
close to half-filling. It leads to a splitting into two bands, and may
lead to an energy gap in the charge excitations spectrum.

In the present paper, which is the continuation and extended version
of our previous work~\cite{hgw}, we investigate the possible phases of
the 2D $t-t'$ Hubbard model on a square lattice with the help of the
free energy expressions derived in Ref.~\cite{gkw} by means of the
flow equations method. In addition to the short paper~\cite{hgw} we
present a summary of the formalism, and give some details
of numerical calculations as well as consider the case of arbitrary
electron concentrations. The phases found by us are discussed in
more details. We find all commonly discussed states as well as a few
new possible phases. We determine in which regions various
symmetry broken phases are more favorable than the symmetric
(i.e. normal) state. This is done for various types of symmetry
breaking independently, that is if no other symmetry breaking would be
present. One phase may suppress another phase. To which extend two
order parameters can coexist with each other is a question, which has
to be investigated in the future. The paper has the following
structure. In Section~\ref{sec2} we summarize the formalism, and
represent the details of numerical calculations. In Section~\ref{sec3}
the numerical results of our stability analysis are given, and the
possible phases of the model are discussed. Section~\ref{sec4} is
devoted to the conclusions.

\section{\label{sec2} Formalism and numerics}

We consider the $t-t'$ Hubbard model 
\begin{eqnarray}
H=\sum_{{\bf k}\sigma}\varepsilon_{\bf k} c^\dagger_{{\bf k}\sigma}
c_{{\bf k}\sigma}
+ {1\over N}\sum_{{\bf k}_1 {\bf k}'_1 \atop {\bf k}_2 {\bf k}'_2}
V({\bf k}_1, {\bf k}_2,{\bf k}'_1,{\bf k}'_2) c^\dagger_{{\bf k}_1\uparrow}
c_{{\bf k}'_1\uparrow} c^\dagger_{{\bf k}_2\downarrow}c_{{\bf k}'_2\downarrow},
\label{ham}
\end{eqnarray}
where $\varepsilon_{\bf k}$ is the Bloch electron energy with the
momentum ${\bf k}$, $c^\dagger_{{\bf k}\sigma} (c_{{\bf k}\sigma})$ is
the creation (annihilation) operator for the electrons with spin
projection $\sigma \in \{\uparrow,\downarrow\}$, and initially $V({\bf
k}_1,{\bf k}_2,{\bf k}'_1,{\bf k}'_2) =U\delta_{{\bf k}_1+{\bf
k}_2, {\bf k}'_1+{\bf k}'_2}$ is the local Coulomb repulsion of two
electrons of opposite spins, $N$ is the number of lattice points,
lattice spacing equals unity.

For a square lattice the single particle dispersion has the form
\begin{eqnarray}
\varepsilon_{\bf k}=-2t(\cos k_x+\cos k_y)-4t'\cos k_x\cos k_y, \label{ek}
\end{eqnarray}
where $t$ is the hopping integral of electrons between nearest
neighbors of the lattice, and $t'$ is the next-nearest-neighbor
hopping integral. The spectrum~(\ref{ek}) contains Van Hove
singularities in the density of states at the energy
$\varepsilon_{VH}=4t'$ related to the saddle points of the Fermi
surface at ${\bf k}=(0,\pm\pi)$ and $(\pm\pi,0)$.  For $t'=0$ and
half-filling the Fermi surface is nested $\varepsilon_{{\bf k}+{\bf
Q}}=-\varepsilon_{\bf k}$ with ${\bf Q}=(\pi,\pi)$, which leads to an
antiferromagnetic instability for $U>0$. The nesting is removed for
$t'/t\neq 0$.

We are mainly interested in the following channels
\begin{eqnarray}
&& V_B({\bf k}, {\bf q})=V({\bf k}, -{\bf k},{\bf q}, -{\bf q}), \\
&&V_H({\bf k}, {\bf q})=V({\bf k}, {\bf q}, {\bf k}, {\bf q}), \\ &&
V_F({\bf k}, {\bf q})=V({\bf k}, {\bf q}, {\bf q}, {\bf k}), \\ &&
V_A({\bf k}, {\bf q})=V({\bf k}, {\bf q}+{\bf Q}, {\bf q}, {\bf
k}+{\bf Q}), \\ && V_C({\bf k}, {\bf q})=V({\bf k}, {\bf q}+{\bf Q},
{\bf k}+{\bf Q}, {\bf q}), \\ && V_Y({\bf k}, {\bf q})=V({\bf k}, {\bf
Q}-{\bf k}, {\bf q}, {\bf Q}-{\bf q}).
\end{eqnarray}
$V_B$ describes the coupling of electron-pairs, $V_H$ and $V_F$ are
interactions showing up in Fermi-liquid theory and describe
(homogeneous) magnetism and Pomeranchuk-instabilities, channels $V_A$
and $V_C$ describe antiferromagnetism and charge density waves with
wave-vector ${\bf Q}$ as well as some other types of ordering. $V_Y$
describes electron pairs with wave-vector ${\bf Q}$. 
These channels are kept within our approach, but finally the effective
interaction $V_Y$ is rather weak. We apply the flow equation
method~\cite{weg,we} to this Hamiltonian, that is we perform a
continuous unitary transformation as a function of a flow parameter
$l$ which brings the Hamiltonian into a mean-field form,
\begin{equation}
\frac{{\rm d}H(l)}{{\rm d}l} = [\eta(l),H(l)], \quad
\eta(l)=[H(l),H^{\rm r}(l)].
\end{equation}
$\eta(l)$ is the generator of the unitary transformation. $H^{\rm
r}(l)$ is obtained from the interaction part of the transformed
Hamiltonian $H(l)$ by
$V^{\rm r}({\bf k}_1, {\bf k}_2,{\bf k}'_1,{\bf k}'_2,l) =r_{\rm eli}({\bf
k}_1, {\bf k}_2,{\bf k}'_1,{\bf k}'_2) V({\bf k}_1, {\bf k}_2,{\bf
k}'_1,{\bf k}'_2,l)$, where $r_{\rm eli}$ is the elimination factor which
indicates how urgently $V({\bf k}_1, {\bf k}_2,{\bf k}'_1,{\bf
k}'_2,l)$ should be eliminated. We choose $r_{\rm eli}({\bf k}_1, {\bf
k}_2,{\bf k}'_1,{\bf k}'_2) =\sum_{\alpha} (v^{\alpha}_{{\bf
k}_1}+v^{\alpha}_{{\bf k}_2} -v^{\alpha}_{{\bf k}'_1}-v^{\alpha}_{{\bf
k}'_2})^2$ with $v^{\alpha}_{\bf k}=-v^{\alpha}_{-\bf
k}=v^{\alpha}_{{\bf k}+\bf Q}$.  This condition guarantees, that those
interactions we wish to keep have $r_{\rm eli}=0$ and are kept, but the
other ones have $r_{\rm eli}>0$ and thus will be eliminated.  In this way
we eliminate the fluctuations around the molecular-field type behavior
and keep finally an interaction for which the molecular-field
treatment is exact.

Keeping these interactions in second order in the Hubbard coupling $U$
by means of the flow equation method~\cite{we} the Hamiltonian is
transformed into an effective one of molecular-field type which
contains products of the biquadratic terms $c^{\dagger}c,\
c^{\dagger}c^{\dagger}$, and $cc$ with total momenta $0$ and ${\bf
Q}$, and depends only on two independent momenta. The second order
effective interactions are $V^{(2)}_B=W_B,\ V^{(2)}_F=W_F,\
V^{(2)}_H=V^{(2)}_F+W_H,\ V^{(2)}_A=W_A,\ V^{(2)}_C=V^{(2)}_A+W_C,\
V^{(2)}_Y=W_Y$, with
\begin{eqnarray}
W_i=-{U^2\over N}\sum_{\bf p_1 p_2} {(\hat{n}_{\bf p_2}-n_{\bf p_1})
(\varepsilon_{\bf p_2}+\hat{\varepsilon}_{\bf p_1}-[\varepsilon_{\bf
a}+ \varepsilon_{\bf b}]/2)\over (\varepsilon_{\bf
p_2}+\hat{\varepsilon}_{\bf p_1}-[\varepsilon_{\bf a}+\varepsilon_{\bf
b}]/2)^2+1/4(\varepsilon_{\bf a}+ \varepsilon_{\bf b})^2}\delta_{{\bf
p_1}+{\bf p_2}, {\bf r}}, \label{ef_in}
\end{eqnarray} where
$i=B, F, H, A, C, Y$, $n_{\bf k}$ is the Fermi distribution function,
$\hat{n}_{\bf p_2}=1-n_{\bf p_2},\ \hat{\varepsilon}_{\bf p_1}=
\varepsilon_{\bf p_1}$ for $i=F,A$, and $\hat{n}_{\bf p_2}=n_{\bf
p_2},\ \hat{\varepsilon}_{\bf p_1}=-\varepsilon_{\bf p_1}$ for
$i=B,H,C,Y$ respectively~\cite{gkw}. The factors $\varepsilon_{\bf a},
\varepsilon_{\bf b}$, and ${\bf r}$ are listed in table~\ref{eab}.
\begin{table}
\caption{\label{eab} The factors $\varepsilon_{\bf a},
\varepsilon_{\bf b}$, and ${\bf r}$ in Eq.~\ref{ef_in}}
\begin{ruledtabular}
\begin{tabular}{cccc}
$i$ & $\varepsilon_{\bf a}$ & $\varepsilon_{\bf b}$ & $\bf r$ \\
\hline
$B$ & $\varepsilon_{\bf k}-\varepsilon_{\bf q}$ & $\varepsilon_{\bf q}-
\varepsilon_{\bf k}$ & ${\bf k}+{\bf q}$ \\
$F$ & $\varepsilon_{\bf k}+\varepsilon_{\bf q}$ & $\varepsilon_{\bf k}+
\varepsilon_{\bf q}$ & ${\bf k}+{\bf q}$ \\
$H$ & $\varepsilon_{\bf k}-\varepsilon_{\bf q}$ & $\varepsilon_{\bf k}-
\varepsilon_{\bf q}$ & ${\bf k}-{\bf q}$ \\
$A$ & $\varepsilon_{\bf k}+\varepsilon_{{\bf q}+{\bf Q}}$ & 
$\varepsilon_{{\bf k}+{\bf Q}}+\varepsilon_{\bf q}$ & 
${\bf k}+{\bf q}+{\bf Q}$ \\
$C$ & $\varepsilon_{\bf k}-\varepsilon_{\bf q}$ & 
$\varepsilon_{{\bf k}+{\bf Q}}-\varepsilon_{{\bf q}+{\bf Q}}$ & 
${\bf k}-{\bf q}$ \\
$Y$ & $\varepsilon_{\bf k}-\varepsilon_{{\bf Q}-{\bf q}}$ & 
$\varepsilon_{\bf q}-\varepsilon_{{\bf Q}-{\bf k}}$ & 
${\bf k}+{\bf q}-{\bf Q}$ \\
\end{tabular}
\end{ruledtabular}
\end{table} 
Then, the free energy has the form
\begin{eqnarray}
\beta F={1\over N}\sum_{\bf k q} \beta U\left(v_1+{U\over t}
V_{{\bf k}, {\bf q}}\right)\Delta_{\bf k}^{\ast}\Delta_{\bf q}+
\sum_{\bf k}f_{\bf k}\Delta_{\bf k}^{\ast}\Delta_{\bf k}=
\sum_{\bf k q}\left( {U\over t}A_{{\bf k}, {\bf q}}+{U^2\over t^2}
B_{{\bf k}, {\bf q}}+\delta_{{\bf k}, {\bf q}}\right)
\sqrt{f_{\bf k}}\Delta_{\bf k}^{\ast}\sqrt{f_{\bf q}}\Delta_{\bf q},
\label{fe}
\end{eqnarray}
where the first term of the middle part of Eq.~(\ref{fe}) is the
energy contribution and the second term is the entropy contribution,
and
\begin{eqnarray}
A_{{\bf k}, {\bf q}}={v_1\beta t\over N \sqrt{f_{\bf k}f_{\bf q}}}, \qquad
B_{{\bf k}, {\bf q}}={\beta tV_{{\bf k}, {\bf q}}\over N
\sqrt{f_{\bf k}f_{\bf q}}},
\end{eqnarray}
with $\beta =1/(k_BT)$, $T$ is the temperature, $V_{{\bf k}, {\bf q}}$
is effective second-order interaction (the factor $U^2/t$ has been
extracted from it), $v_1=\pm 1$ is the sign with which the first order
contribution enters into the effective interaction, $f_{\bf k}$ is an
entropy coefficient, and $\Delta_{\bf k}$ are the order
parameters. For example, $\langle c_{{\bf k}\sigma} c_{{\bf
-k}\sigma'}\rangle = (\sigma_y)_{\sigma\sigma'}\Delta_{\bf k}^{s}+
\sum_{\alpha}(\sigma_y\sigma_{\alpha})_{\sigma\sigma'} \Delta_{\bf
k}^{t\alpha}$, where $\sigma_{\alpha}$ is a Pauli spin matrix
($\alpha=x, y, z$), and $\Delta_{\bf k}^{s}\ (\Delta_{\bf
k}^{t\alpha})$ is the singlet (triplet) amplitude. An expression
similar to Eq.~(\ref{fe}) is obtained for particle-hole channels with
the order parameters $\nu$ instead of $\Delta$. In this case, for
example, we have $\langle c^\dagger_{{\bf k}\sigma}c_{{\bf k}+{\bf
Q}\sigma'}\rangle = \nu_{\bf
k}^{s}\delta_{\sigma,\sigma'}+\sum_{\alpha}\nu_{\bf k}^{t\alpha}
(\sigma_{\alpha})_{\sigma\sigma'}$.

In total we consider four channels, two particle-hole and two
particle-particle ones, since the total momentum can be 0 and ${\bf
Q}$. In all cases we distinguish between singlet and triplet
excitations, and in the channels corresponding to the total momentum
${\bf Q}$ we distinguish also between $\Delta_{{\bf k}+{\bf
Q}}=\pm\Delta_{\bf k}$ and $\nu_{{\bf k}+{\bf Q}}=\pm\nu_{\bf k}$
respectively. Therefore, we introduce the following notations: $ph$
denotes particle-hole channel, $si\ (tr)$ denotes singlet (triplet),
$q_{\pm}$ corresponds to $\Delta_{{\bf k}+{\bf Q}}=\pm\Delta_{\bf k}$
and $\nu_{{\bf k}+{\bf Q}}=\pm\nu_{\bf k}$ respectively, otherwise
$0$.

In Eq.~(\ref{fe}) and the analogous expressions for particle-hole channels one 
has
\begin{equation}
\begin{array}{cccc}
{\rm channel} & v_1 & V_{{\bf k},{\bf q}} & f_{\bf k} \\ \hline
pp\ si/tr\ 0 & +1 & V_B^{(2)}({\bf k},{\bf q}) &
f(\beta(\varepsilon_{\bf k}-\mu),\beta(\mu -\varepsilon_{\bf k})) \\
ph\ si\ 0 & +1 & 2V_H^{(2)}({\bf k},{\bf q})-V_F^{(2)}({\bf k},{\bf q}) &
f(\beta(\varepsilon_{\bf k}-\mu),\beta(\varepsilon_{\bf k}-\mu)) \\
ph\ tr\ 0 & -1 & -V_F^{(2)}({\bf k},{\bf q}) &
f(\beta(\varepsilon_{\bf k}-\mu),\beta(\varepsilon_{\bf k}-\mu)) \\
pp\ si/tr\ q_{\pm} & +1 & V_Y^{(2)}({\bf k},{\bf q}) &
f(\beta(\varepsilon_{\bf k}-\mu),\beta(\mu -\varepsilon_{{\bf k}+{\bf Q}})) \\
ph\ si\ q_{\pm} & +1 & 2V_C^{(2)}({\bf k},{\bf q})-V_A^{(2)}({\bf k},{\bf q}) &
f(\beta(\varepsilon_{\bf k}-\mu),\beta(\varepsilon_{{\bf k}+{\bf Q}}-\mu)) \\
ph\ tr\ q_{\pm} & -1 & -V_A^{(2)}({\bf k},{\bf q}) &
f(\beta(\varepsilon_{\bf k}-\mu),\beta(\varepsilon_{{\bf k}+{\bf Q}}-\mu)).
\end{array}
\end{equation}
 
The entropy coefficient function is
\begin{eqnarray}
f(x,y)={x-y\over e^x-e^y}(e^x+1)(e^y+1).
\end{eqnarray}

We start from the symmetric state and investigate whether this state
is stable against fluctuations of the order parameters $\Delta$ and
$\nu$. As soon as a non-zero $\Delta$ or $\nu$ yields a lower free
energy in comparison with the symmetric state with all vanishing
$\Delta$ and $\nu$, then the symmetric state is unstable and the
system will approach a symmetry broken state. This indicates a phase
transition. Of course, one phase may suppress another phase. To which
extend two order parameters can coexist with each other is a question,
which has to be investigated. But this problem lies outside the scope
of the present paper.

We performed numerical calculation on a square lattice with $24\times
24$ points in the Brillouin zone for the various representations under
the point group $C_{4\nu}=4mm$. This group consists of the following
symmetry elements: mirror reflections with respect to the $x$- and
$y$-axes and the lines $y=\pm x$, a fourfold rotation about the axis
which is perpendicular to $x$-$y$-plane. The representations of the
even-parity states are one-dimensional. We denote them by $s_+=s_1,\
s_-=g=s_{xy(x^2-y^2)},\ d_+=d_{x^2-y^2},\ d_-=d_{xy}$. The odd-parity
representation is two-dimensional, here simply denoted by $p$.

Initially, such numerical calculations have been performed in
Refs.~\cite{gkw,gr}, but they were sensitive to the lattice size at
low temperatures, specially for the particle-hole channels. The main
problem is that the wave-vector space contributing essentially to the
order parameter is very small at low temperatures as a result of
exponential divergence of the entropy coefficients $f_{\bf k}$. As we
consider the stability of the symmetric state with respect to a symmetry
broken phase (thus we are interested in small fluctuations of free
energy~(\ref{fe}) around zero) and the energy contribution to
Eq.~(\ref{fe}) changes slowly, the entropy contribution has to change
slowly also. Rewriting Eq.~(\ref{fe}) in terms of new renormalized
order parameter $\tilde{\Delta}_{\bf k}=\sqrt{f_{\bf k}}\Delta_{\bf
k}$, we conclude that the averaging $\langle {1\over f_{\bf
k}}\rangle$ has to be performed in ${\bf k}$-space when we are
calculating entropy coefficients. For the value of the inverse entropy
coefficient $1/f_{\bf k_0}$ we take the average value of the
coefficients over $N_a\times N_a$ lattice points of a grid with center
at the ${\bf k_0}$-point and the spacing $2\pi/(N_a\sqrt{N})$. The
number $N_a$ of points in momentum space is inversely proportional to
temperature and $\sqrt{N}$, all together these points cover the
Brillouin zone.  A similar procedure is applied to the chemical
potential calculation. One should also perform the averaging in
determination of effective interactions, but it would take much more
time.  We have checked the size effects in the framework of such an
improved scheme, performing some calculations also for $16\times 16$
and $32\times 32$ lattices, and found that the differences are very
small and unessential. Although size effects increase at low
temperatures in the low density region, they do not touch and change
the leading instabilities essentially.

For each channel the eigenvalues of the matrix ${U\over t}A+{U^2\over
t^2}B$ have to be determined. Whenever the lowest eigenvalue $\lambda$
(i.e. the most negative) equals -1, then a critical $T_c/t$ or
$(U/t)_c$ is reached. For the $s_+$ representation (in case of the
effective interactions $V_A,\ V_C,\ V_Y$ only for $\nu_{{\bf k}+{\bf
Q}}=+\nu_{\bf k},\ \Delta_{{\bf k}+{\bf Q}}=+\Delta_{\bf k}$) one has
to find the solution by iterating the eigenvalue equation as a
function of $U/t$. In the case of all other representations the
$A$-term does not contribute. Therefore then we determine the lowest
eigenvalue $\lambda$ of the matrix $B$, and obtain $(U/t)_c=
1/\sqrt{-\lambda}$.

\section{\label{sec3} Possible phases of the model}
			
\begin{figure}
\includegraphics{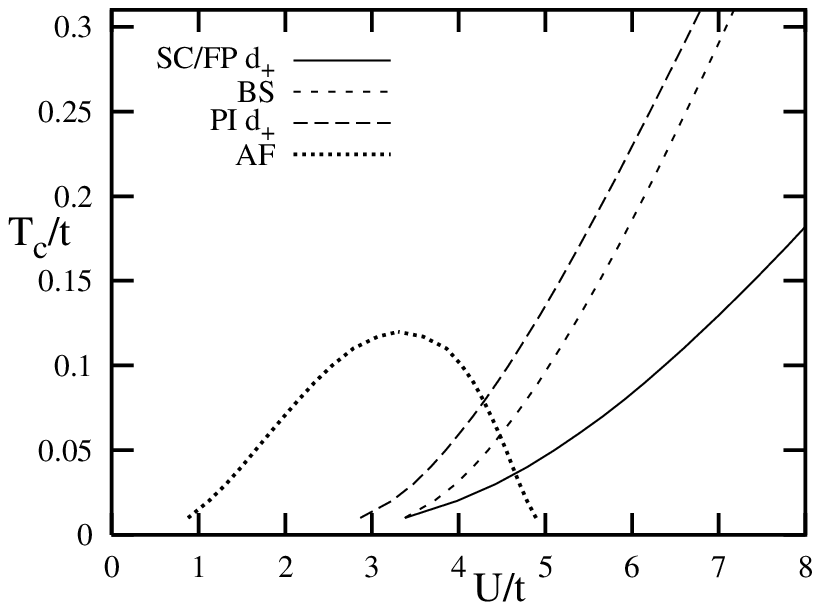}
\caption{\label{fig1} Temperature phase diagram of the 2D $t-t'$
Hubbard model for $n=1,\ t'=0$. Chemical potential $\mu=0$. SC stands
for superconductivity, FP for flux phase, BS for band splitting, PI
for Pomeranchuk instability, AF for antiferromagnetism.}
\end{figure}

\begin{figure}
\includegraphics{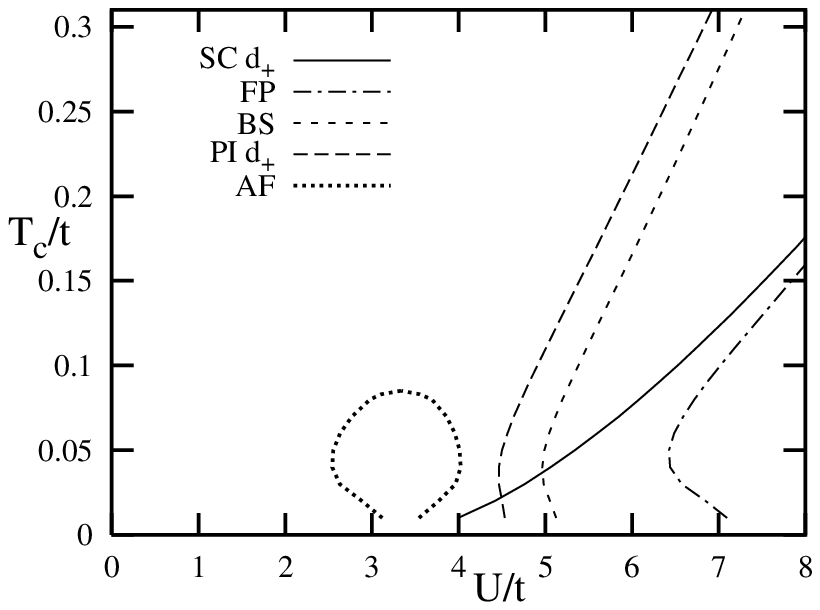}
\caption{\label{fig2} Temperature phase diagram of the 2D $t-t'$
  Hubbard model for $n=0.94,\ t'=0$. Chemical potential varies between
  $\mu/t=-(0.097\div 0.146)$. Notations are the same as in
  Fig.~\ref{fig1}.}
\end{figure}

We start from $t'=0$ and half-filling $n=1$ (see Fig.~\ref{fig1}).  As
expected in this case the leading instability is the antiferromagnetic
one, it appears in the $ph\ tr$ channel with $q_+$ symmetry and $s_+$ wave
character. As we leave half-filling the antiferromagnetism becomes
weaker (Fig.~\ref{fig2}) and disappears at hole doping $\delta\equiv
1-n \sim 0.08$. As soon as the system is doped the nesting is reduced, since there are
no longer points on the Fermi surface connected by the wave-vector ${\bf Q}$ and
we observe a reentrant behavior. 
The second order contribution to the antiferromagnetic
channel suppresses antiferromagnetism (at least in the
$s$-channel). Therefore antiferromagnetism disappears at larger values
of $U/t$. Strong coupling calculations yield a $t-J$-model in which the
exchange coupling decreases like $J=4t^2/U$ for large $U$. Thus, our
observation is remarkable, since we work with a weak-coupling
calculation, we obtain at intermediate couplings the same tendency as
it is expected at strong interactions although we do not reproduce the
Ne\'{e}l temperature behavior $T_c\sim t^2/U$.  We take the decrease
of the Ne\'{e}l temperature as an indication that the calculation in
second order in $U$ yields reasonable results even for intermediate
values $U\sim 4t$. For stronger couplings higher order contributions
will become important. However, it is of interest to see which
instabilities emerge within our approximation for larger values of
$U/t$, since this gives a hint which types of ordering should be
investigated for stronger couplings. Therefore, we discuss the phase
diagram obtained from the second order calculation also for larger
values $U/t$.  It might be also that higher order contributions
neglected by us enhance the tendency towards antiferromagnetism, and
then it will decay not so rapidly. Nevertheless, we emphasize that
already second order calculations give the qualitative description of
antiferromagnetism at intermediate couplings.  In the latter context,
we make a comparison of our results with experimental data on some
high-temperature cuprate superconductors and transition metal
compounds. As one can see from Fig.~\ref{fig1} antiferromagnetism
disappears at the temperature $T\approx 0.1t$. Taking the estimation
for the hopping integral $t\approx 0.1\div 0.3$~eV, which is the
realistic values for transition metal and their
compounds~\cite{he,fr}, we obtain the N\' eel temperatures
$T_c=T_N\approx 110\div 350$~K. These values agree with experimental
data on half-filled antiferromagnetic materials~\cite{mot}. Let us
consider also the parent compound La$_2$CuO$_4$ of high-$T_c$ cuprate
systems La$_{2-x}M_x$CuO$_4$ ($M=$~Sr, Ba, Ca). It is an
antiferromagnetic insulator with the N\' eel temperature $T_N\sim
300$~K and half-filled $d$-band~\cite{ift}. The electronic structure
of these systems is considered to be described by the three-band
Hubbard model (or $d-p$ model)~\cite{pf,ift}. Some of the model
parameters deduced from photoemission and LDA calculations are
$U_{dd}\approx 6\div 10$~eV, and $t_{pd}\approx 1\div
1.5$~eV~\cite{ift}. The main features of the three-band Hubbard model
can be mapped onto the single-band Hubbard model, and therefore one
may describe~\cite{sc} these compounds by the Hubbard
model~(\ref{ham}). The values of the parameters $U$ and $t$ of such a
single-band Hubbard model are smaller in comparison with the
corresponding parameters $U_{dd}$ and $t_{pd}$ of the three-band
extended Hubbard model, the Coulomb repulsion $U$ is specially
reduced~\cite{we1}. Then, taking the values $t\approx 1$~eV and
$U/t\approx 4.7\div 4.8$ (it corresponds to $T_c/t\approx 0.03\div
0.02$) we obtain the theoretical estimation of the N\' eel temperature
for La$_2$CuO$_4$ around $T_N\approx 290$~K, that is very close to the
experimental value.

\begin{figure}
\includegraphics{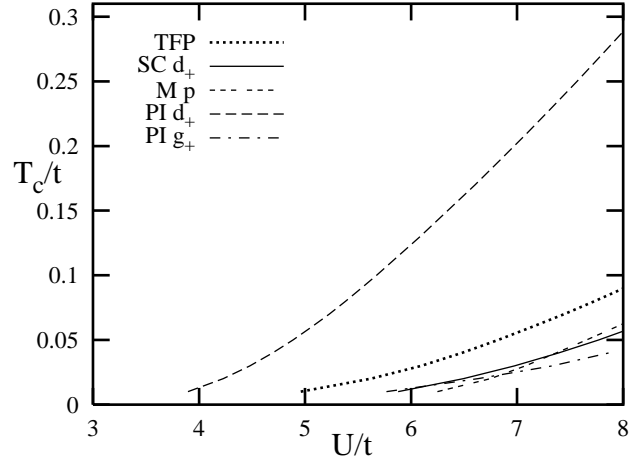}
\caption{\label{fig3} Temperature phase diagram of the 2D $t-t'$
Hubbard model for $t'=-t/3$, and $n=0.68$ (the Van Hove
filling). Chemical potential varies between $\mu/t=-(1.317\div
1.339)$. TFP stands for triplet flux phase, M for magnetic
particle-hole instability in triplet channel, other notations are the
same as in Fig.~\ref{fig1}.}
\end{figure}

\begin{figure}
\includegraphics{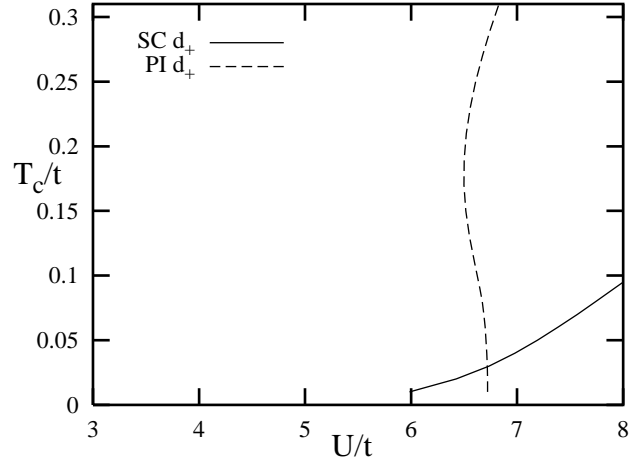}
\caption{\label{fig4} Temperature phase diagram of the 2D $t-t'$
Hubbard model for $t'=-5t/12$, and $n=0.90$. Chemical potential varies
between $\mu/t=-(1.109\div 0.985)$. Notations are the same as in
Fig.~\ref{fig3}.}
\end{figure}

\begin{figure}
\includegraphics{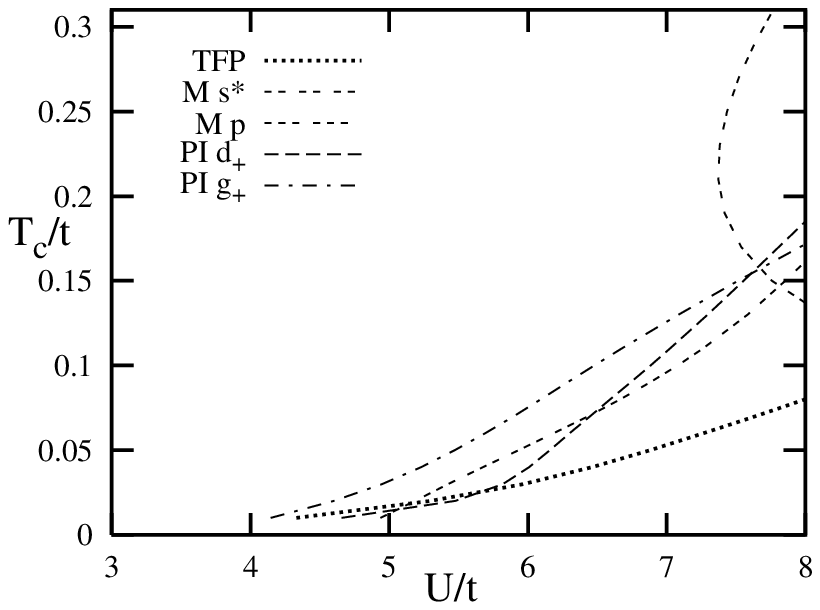}
\caption{\label{fig5} Temperature phase diagram of the 2D $t-t'$
Hubbard model for $t'=-5t/12$, and $n=0.55$ (the Van Hove
filling). Chemical potential varies between $\mu/t=-(1.666\div
1.632)$. Notations are the same as in Fig.~\ref{fig3}.}
\end{figure}

\begin{figure}
\includegraphics{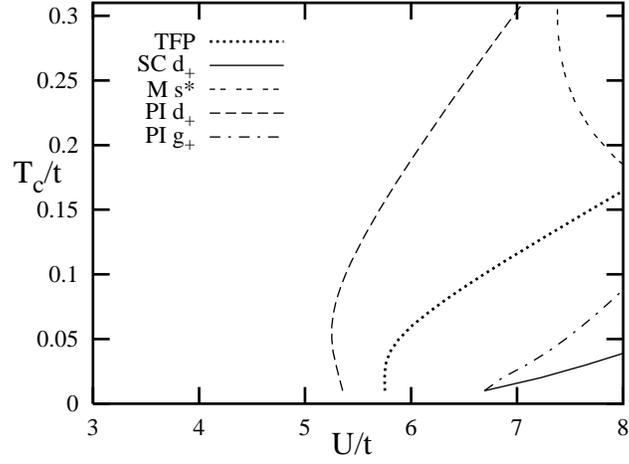}
\caption{\label{fig6} Temperature phase diagram of the 2D $t-t'$
Hubbard model for $t'=-5t/12$, and $n=0.70$ (above the Van Hove
filling). Chemical potential varies between $\mu/t=-(1.493\div
1.380)$. Notations are the same as in Fig.~\ref{fig3}.}
\end{figure}

The next instability is a Pomeranchuk instability with
$d_{x^2-y^2}$-wave symmetry in the singlet channel (namely $ph\ si\
0$). The corresponding eigenvector signals a deformation of the Fermi
surface which breaks the point group symmetry of the square
lattice. At high temperatures the system has a tetragonal structure
and an orthorhombic one at low temperatures. The $d_{x^2-y^2}$ wave
Pomeranchuk instability dominates in the region of temperature
$T>0.05t$ at small hole doping ($\delta <0.2$), and for negative
$t'\geq -t/3$ the Pomeranchuk instability dominates in all temperature
region at the electron concentrations around the Van Hove filling (see
Figs.~\ref{fig3},~\ref{fig4}). It competes with other instabilities at
$t'<-t/3$ and the electron concentrations around the Van Hove filling,
and it is not the leading one here (Fig.~\ref{fig5}). But the $d$-wave
Pomeranchuk instability appears again to dominate at large negative
$t'<-t/3$ in the region of electron density above the Van Hove filling
and below some doping (for example, see Fig.~\ref{fig6}); for the
value $t'=-5t/12$ this value of doping equals $\delta\approx 0.2$. In
agreement with the ideas of Ref.~\cite{hm} the instability is mainly
driven by a strong attractive interaction between particles on
opposite corners of the Fermi surface near the saddle points and a
repulsive interaction between particles on neighboring corners. To
favor such a behavior we need a sizable $t'$ reducing
antiferromagnetic correlations.  The critical temperature of the
$d$-wave Pomeranchuk instability occurrence decreases with increasing
hole doping, and at sufficiently large values of doping and away from
Van Hove filling this phase requires strong couplings. Thus the hole
doping suppresses the tendency towards an orthorhombic distortion of
the Fermi surface (or lattice). The tendency towards a spontaneous
deformation of Fermi surface with $d$-wave character has been observed
by A.~P.~Kampf and A.~A.~Katanin~\cite{kk} to be one of the leading
instabilities in the 2D extended $U$-$V$-$J$ Hubbard model in the
parameter range which is relevant for underdoped cuprate systems. A
similar Pomeranchuk instability
has been found by the authors of Ref.~\cite{okf} to be a
characteristic of a nematic Fermi fluid~\cite{kfe} in two dimensions.

\begin{figure}
\includegraphics{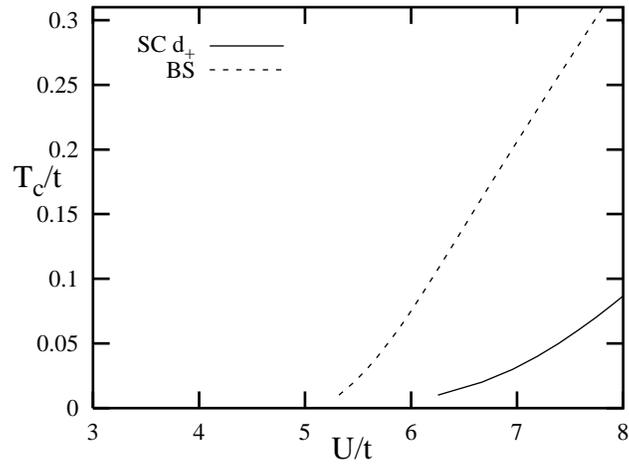}
\caption{\label{fig7} Temperature phase diagram of the 2D $t-t'$
Hubbard model for $t'=-t/6$, and $n=1.15$. Chemical potential varies
between $\mu/t=-(0.001\div 0.080)$. Notations are the same as in
Fig.~\ref{fig3}.}
\end{figure}

Another possible phase is a particle-hole instability of singlet type
with staggered $p$-wave symmetry ($ph\ si\ q_-\ p$). The corresponding
quasiparticle energy spectrum is
\begin{eqnarray}
E_{\bf k}=\pm\sqrt{\varepsilon^2_{\bf k}+\tilde{\varepsilon}_{\bf k}^2}
\end{eqnarray}
where 
\begin{eqnarray}
\tilde{\varepsilon}_{\bf k}={1\over N}\sum_{\bf q}[2V_C({\bf k},{\bf q})-
V_A({\bf k},{\bf q})-2V_C({\bf k},{\bf q}+{\bf Q})+
V_A({\bf k},{\bf q}+{\bf Q})]\nu_{\bf q}^s
\end{eqnarray}
varies like $\tilde{\varepsilon}_x\sin k_x+\tilde{\varepsilon}_y\sin
k_y$ with roughly constant $\tilde{\varepsilon}_x$ and
$\tilde{\varepsilon}_y$.  It yields a splitting into two bands and may
lead to an energy gap in the charge excitations spectrum. Another
mechanism for a charge gap formation has been proposed~\cite{mj,klpt}
recently in the 2D Hubbard model with $t'=0$ at weak coupling. The
band splitting phase is developed in the region of electron
concentration around half-filling, and is one of the strongest in that
region. This instability dominates when the electron concentration
$n>1.10$ and $t'\neq 0$ (Fig.~\ref{fig7}). The negative
next-nearest-neighbors hopping favors this phase to be dominant
instability at electron doping, but the value of coupling strength
$U/t$ required for the occurrence of this instability increases with
increasing $|t'|$. The critical temperature of the transition to the
band splitting phase decreases with increasing electron doping, thus
the electron doping suppresses the tendency towards this phase.

The singlet superconducting $d_{x^2-y^2}$ instability ($pp\ si\ 0\
d_+$) coincides with the $d_{x^2-y^2}$-wave staggered flux phase at
half-filling and $t'=0$ (the discussion on flux phases appears in the
next paragraph). Away from half-filling the degeneration disappears,
and $d$-wave superconductivity dominates at low temperatures in
certain regions of electron concentration around half-filling which
depend on the value of $t'\neq 0$. Even large values of $|t'|$ do not
destroy the dominant low-temperature behavior of $d_{x^2-y^2}$-wave
superconductivity at hole doping (Fig.~\ref{fig4}). It is not
destroyed also at sufficiently large hole doping: for small $t'\geq
-t/6$ even at the values of hole doping $\delta\sim 0.3$, and for
$t'\leq -t/3$ at $\delta\sim 0.15$ the $d$-wave superconductivity
dominates at low temperatures. However, this dominant behavior is
destroyed around the Van Hove filling where other phases appear to
dominate. At small electron doping $d$-wave superconductivity
dominates at low temperatures also, but it is not the leading
instability at $n-1\geq 0.1$ and $t'\neq 0$ (see Fig.~\ref{fig7}). In
contrast to the pure 2D Hubbard model with $t'=0$, an electron-hole
asymmetry of the $d$-wave superconducting phase is manifested in the
$t-t'$ Hubbard model. In this model with negative $t'$ the $d$-wave
superconductivity appears to dominate within wider region of hole
doping in comparison with the electron doping in agreement with the
one-loop renormalization group analysis~\cite{ho}. In this connection it
should be mentioned that the most examples of discovered
high-temperature superconducting materials are hole-doped systems, and
only a few examples are electron-doped ones.

\begin{figure}
\includegraphics{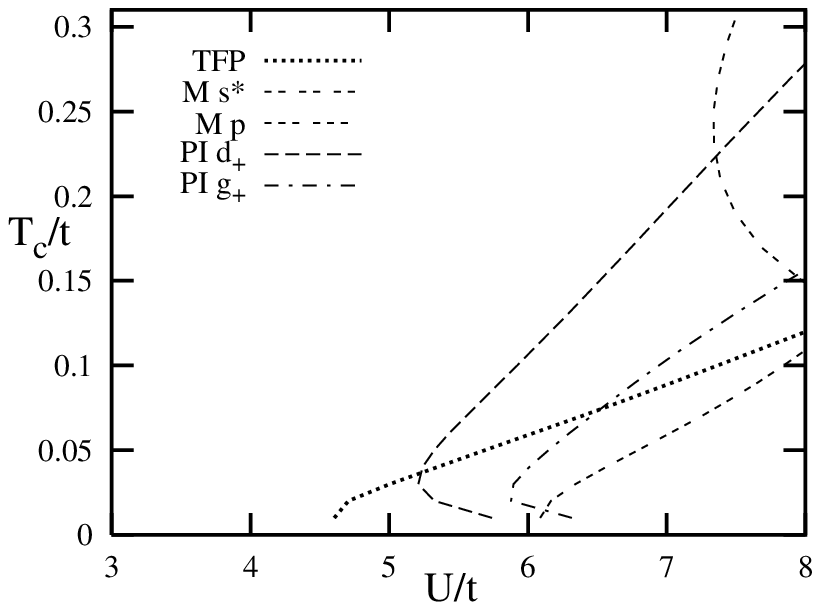}
\caption{\label{fig8} Temperature phase diagram of the 2D $t-t'$
Hubbard model for $t'=-5t/12$, and $n=0.60$ (slightly above the Van
Hove filling). Chemical potential varies between $\mu/t=-(1.621\div
1.550)$. Notations are the same as in Fig.~\ref{fig3}.}
\end{figure}

We observe also particle-hole instabilities with staggered symmetry of
$d_+$ wave character in singlet and triplet channels ($ph\ si/tr\ q_-\
d_+$). These states are singlet and triplet flux phases which are
approximately described by the operators structures
\begin{eqnarray}
&&i\sum_{{\bf k}\sigma}(\cos k_x-\cos k_y)c^\dagger_{{\bf k}+{\bf Q}\sigma}
c_{{\bf k}\sigma}, \label{sfp}
\\
&&i\sum_{{\bf k}\sigma\sigma '}(\cos k_x-\cos k_y)\vec{\sigma}_{\sigma\sigma '}
c^\dagger_{{\bf k}+{\bf Q}\sigma}c_{{\bf k}\sigma '}, \label{tfp}
\end{eqnarray}
respectively, where the components of $\vec{\sigma}$ are $2\times 2$
spin Pauli matrices. For $t'=0$ critical temperatures of the
transition to the singlet and triplet flux phases are degenerate, in
this case these phases vanish at hole doping $\delta\sim 0.12$ in
agreement with a mean-field solution of the $t-J$ model presented in
Ref.~\cite{hma}. For $t'\neq 0$ they occur more easily at the electron
concentrations which are slightly above the Van Hove filling, but the
singlet flux phase disappears for $t'\leq -t/3$.  For non-zero values
of the next-nearest-neighbors hopping the critical transition
temperatures of these phases are different, and the triplet one is
higher. Moreover, the triplet analog of flux phase dominates at low
temperatures and $t'=-5t/12$ when the electron concentration is
slightly above the Van Hove filling (see Fig.~\ref{fig8}) in contrast
to the results of Ref.~\cite{hs} which point out the occurrence of
triplet superconductivity with $p$-wave symmetry in this region. The
triplet flux phase is also one of the leading instabilities for
various values of next-nearest-neighbors hopping $t'$ and certain
region of electron concentrations (for example, see
Fig.~\ref{fig3},~\ref{fig5},~\ref{fig6}).

The singlet flux state, described by Eq.~(\ref{sfp}), breaks
time-reversal, translational, and rotational symmetries. This state is
a phase of circular charge currents flowing around the plaquettes of a
square lattice with alternating directions.  The authors of
Ref.~\cite{hma} found that these currents generate a small magnetic
field, also forming a staggered pattern, and they estimated the field
strength just above the plaquette center to be around 10~G. They
pointed out that muon-spin-rotation experiments may be able to detect
this magnetic field. Recently, the idea of circulating orbital
currents (also called an orbital antiferromagnetism) has been
discussed as promising candidates for a description of the pseudogap
phase of high-$T_c$ cuprates~\cite{clmn,ckn,le}. There is also a
proposal~\cite{va} of a circulating current phase, which does not
break translational symmetry. The possible existence of orbital charge
currents has most recently received experimental
support. Angle-resolved photoemission with circularly polarized light
identified intensity differences for left- and right-circularly
polarized photons in the psedogap phase of
Bi$_2$Sr$_2$CaCu$_2$O$_{8+\delta}$~\cite{krf}. Also small $c$-axis
oriented ordered magnetic moments, and an evidence for static
alternating magnetic fields below the psedogap temperature in the
underdoped YBa$_2$Cu$_3$O$_{6+x}$ systems were observed by neutron
scattering~\cite{mo} and muon-spin-rotation~\cite{mil} measurements
respectively. These data can find a natural interpretation in terms of
planar circulating current phases~\cite{tde}.

The triplet flux phase, described by Eq.~(\ref{tfp}), breaks
translational, rotational and spin-rotational symmetries, and does not
break time-reversal invariance. This phase is a state in which spin-up
and spin-down electron currents circulate around the plaquettes in
opposite directions to produce non-zero spin currents. It corresponds
to an alternating pattern of spin currents. Similar triplet flux phase
(so called spin flux phase or spin nematic state) has been discussed
by Narsesyan and co-workers~\cite{njk} by means of a mean-field
theory, and more recently in Refs.~\cite{bbd,kk} by means of
renormalization group approaches in the framework of the 2D extended
$U$-$V$-$J$ Hubbard model. Another triplet flux phase, which breaks
time-reversal invariance, has been considered by Nayak~\cite{na} as a
triplet analog of the density wave order parameter potentially
relevant to the cuprates. To our knowledge a triplet version of the
flux phase has not yet been observed in numerical solutions of the 2D
$t-t'$ Hubbard model. In the case of polarized particle-hole pairs (it
means that we take $\sigma_z$ instead of $\vec{\sigma}$ in
Eq.~(\ref{tfp})) the alternating pattern of spin currents will
generate electric field, which could be, in principle, measurable in
experiments. As triplet flux phase~(\ref{tfp}) does not have anomalous
expectation values for the spin density but, rather, for spin
currents, these excitations do not contribute to the spin-spin
correlation function, therefore spin currents do not couple to
photons, neutrons or nuclear spins~\cite{na}. However, they could be
detected with nuclear quadrupole resonance or two-magnon Raman
scattering experiments~\cite{na}.

\begin{figure}
\includegraphics{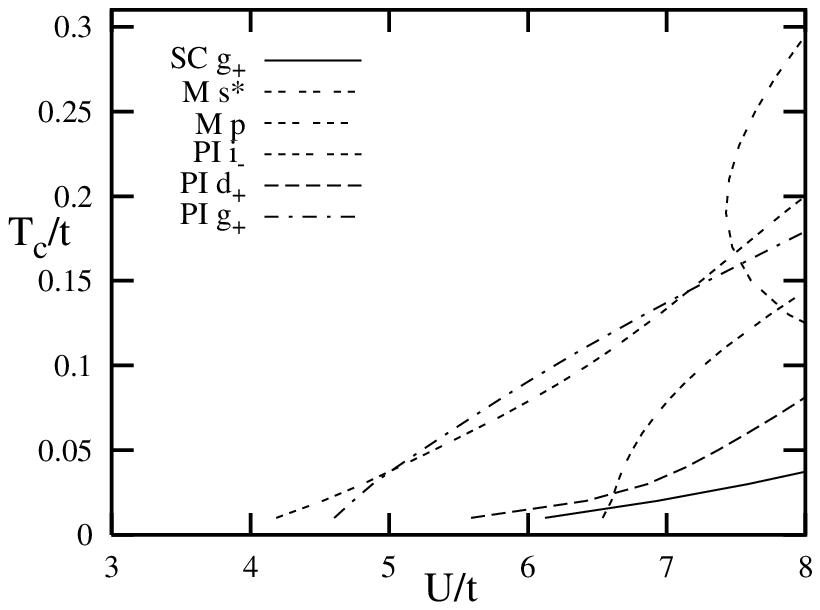}
\caption{\label{fig9} Temperature phase diagram of the 2D $t-t'$
Hubbard model for $t'=-5t/12$, and $n=0.50$ (slightly below the Van
Hove filling). Chemical potential varies between $\mu/t=-(1.709\div
1.713)$. Notations are the same as in Fig.~\ref{fig3}.}
\end{figure}

\begin{figure}
\includegraphics{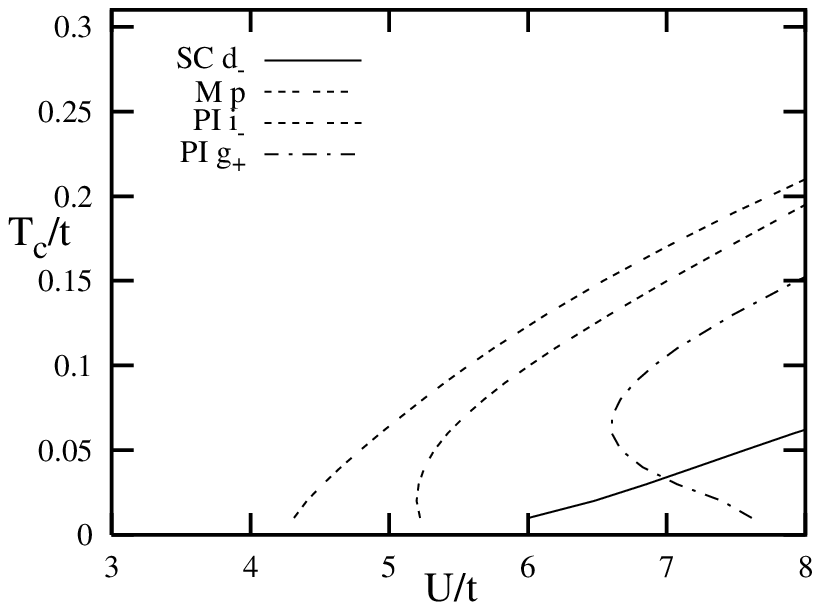}
\caption{\label{fig10} Temperature phase diagram of the 2D $t-t'$
Hubbard model for $t'=-5t/12$, and $n=0.35$ (below the Van Hove
filling). Chemical potential varies between $\mu/t=-(1.885\div
1.963)$. Notations are the same as in Fig.~\ref{fig3}.}
\end{figure}

At $t'=-5t/12$ a few other new instabilities appear to compete at the
Van Hove filling and low temperatures (Fig.~\ref{fig5}) in
disagreement with the conclusions of Ref.~\cite{hs} on the occurrence
of ferromagnetism. The leading one is another Pomeranchuk instability
in the $s_+$ channel with $g_+=g_{x^4+y^4-6x^2y^2}$ wave character
($ph\ si\ 0\ s_+$, 4 node lines in ${\bf k}$-space).  This phase occurs
more easily when the electron concentration is close to the Van Hove
filling, and it is a leading instability at the Van Hove filling and
the values of electron concentration which are slightly smaller than
the Van Hove filling (Figs.~\ref{fig5},~\ref{fig9}). The $g_+$
Pomeranchuk instability appears at sizable values of
next-nearest-neighbors hopping $t'\sim -t/3$ (Fig.~\ref{fig3}), but it
requests sufficiently large absolute values of $t'$ ($t'<-t/3$) to be
one of the dominant instabilities.  In the $d_-$ channel an $i$-wave
(6 node lines in ${\bf k}$-space) Pomeranchuk instability ($ph\ si\ 0\
d_-$) appears when electron concentration $n$ is smaller than the Van
Hove filling at $t'=-5t/12$ (Fig.~\ref{fig9}). It is a leading one at
small values of the electron concentration (Fig.~\ref{fig10}), and
dominates in the temperature region $T>0.15t$ at $t=-t'/3$ when the
electron concentration is sufficiently below the Van Hove filling. The
increase of electron concentration reduces essentially this phase.

When the electron concentration is decreased below the Van Hove
density at $t'=-5t/12$, a particle-hole instability of $p$-wave
symmetry in triplet channel ($ph\ tr\ 0\ p$) dominates (see
Fig.~\ref{fig10}). The electron operators structure which describes
this magnetic phase is roughly
\begin{eqnarray}
i\sum_{{\bf k}\sigma\sigma '}(t_x^{M}\sin k_x +t_y^{M}\sin
k_y)\vec{\sigma}_{\sigma\sigma '}c^\dagger_{{\bf k}\sigma}c_{{\bf
k}\sigma '}= {i\over 2}\sum_{{\bf R}\sigma\sigma '}\left(t_x^{M}
[c^\dagger_{{\bf R}+{\bf e}_x,\sigma}-c^\dagger_{{\bf R}-{\bf
e}_x,\sigma}] +t_y^{M}[c^\dagger_{{\bf R}+{\bf
e}_y,\sigma}-c^\dagger_{{\bf R}-{\bf
e}_y,\sigma}]\right)\vec{\sigma}_{\sigma\sigma '}c_{{\bf R},\sigma '}
\label{mp}
\end{eqnarray}
with ${\bf e}_x$ (${\bf e}_y$) being the unit vector along the $x$-
($y$-) axis, ${\bf R}$ denotes lattice site. The quantity $t^M$ is
given by $t_x^{M}\sin k_x +t_y^{M}\sin k_y \approx -1/N\sum_{\bf q}
V_F({\bf k},{\bf q})\nu_{\bf q}^{t}$. One can see from Eq.~(\ref{mp})
that this instability gives rise to a phase of magnetic currents, the
magnetization equals zero as a result of $p$-wave character of the
instability. The magnetic currents phase dominates also at very low
temperatures and electron concentrations which are slightly smaller
the Van Hove filling for $t'=-5t/12$ (Fig.~\ref{fig9}), and in the
temperature region $T<0.15t$ below the Van Hove density for $t'=-t/3$.

In the temperature region $T>0.15t$ a particle-hole instability with
$s^*$-wave character ($ph\ tr\ 0\ s_+$, its order parameter changes
sign close to the Fermi-edge) in the triplet $s_+$ channel dominates
at $t'\leq -5t/12$ and the density range around the Van Hove filling
(see Figs.~\ref{fig5},~\ref{fig8},~\ref{fig9}). It is likely that the
order parameter contributions do not compensate exactly, so that a
weak ferromagnetism appears.  When the electron concentration is
increased above the Van Hove filling this instability does not become
weaker, but the $d_{x^2-y^2}$ wave Pomeranchuk and the triplet flux
phase instabilities are stronger and dominate at low
temperatures. Then, this $s^*$-magnetic phase disappears at
sufficiently large values of electron concentration in comparison with
the Van Hove filling, or smaller $|t'|$. With the decrease of electron
concentration from the Van Hove filling this instability becomes
weaker and disappears to be a leading one at small electron density
(Fig.~\ref{fig10}).  From Figs.~\ref{fig5},~\ref{fig8},~\ref{fig9} one
can see a reentrant behavior of the $s^*$-magnetic phase in some
region of the values $U/t$: approaching $T_c$ we get $T_c^{l}$ from
low temperatures and $T_c^{u}$ from high temperatures at the same
value of coupling $U/t$ ($T_c^{l}\neq T_c^{u}$). This is a result of
different behavior of $T_c(U/t)$ in two regimes. First regime occurs
in the region where the $s^*$-magnetic instability dominates and the
transition from a paramagnetic state to the $s^*$-magnetic phase
occurs directly without any intermediate phase, it corresponds to the
temperatures $T>0.15t$ in Figs.~\ref{fig5},~\ref{fig9}. In this case
the critical temperature increases with the increase of correlation
strength $U/t$. Therefore electron correlations enhance the tendency
towards the $s^*$-magnetic instability.  Another regime occurs at the
temperatures $T<0.15t$, where the $s^*$-magnetic phase is not a
leading instability and other instabilities dominate.  In this
situation the critical temperature exhibits an anomalous behavior, it
decreases with increasing the coupling $U/t$. The $s^*$-magnetic phase
is reduced.  Since at lower temperatures only a smaller region in ${\bf
k}$-space around the Fermi-edge contributes, the sign-change of the
order parameter reduces the effective interaction. As the transition
to this phase requires some finite value of the effective interaction
at low temperatures, the correlation strength has to increase rapidly.

We observe also a few other weaker instabilities. Apart from
$d_{x^2-y^2}$ wave superconductivity mentioned above other two
superconducting instabilities appear. At $t'\leq -t/3$ a $d_{xy}$ wave
superconducting phase occurs in the singlet channel sufficiently below
the Van Hove filling (see, for example, Fig.~\ref{fig10}), and for
larger absolute values $t'\leq -5t/12$ we observe a $g_+$ wave
superconductivity in the $s_+$ singlet channel below the Van Hove
filling (Fig.~\ref{fig9}). All these phases require strong couplings.

The peculiar feature of the superconducting phase in comparison with
the other phases observed by us should be noted. Away from the Van
Hove filling when temperature approaches zero the curves corresponding
to superconducting phase are flat, whereas the curves corresponding to
all other phases observed become steep. Therefore, at very low
temperatures the transition from a paramagnetic phase to the
superconducting one can occur at very small values of the
corresponding effective interaction in contrast with the transitions
to other possible phases which require some finite values of the
effective interactions. One can see also that the critical
temperatures of all phases (excepting antiferromagnetism and
$s^*$-magnetic phase in certain regions) increase with the increase of
correlation strength $U/t$. Thus, electron correlations enhance the
tendency towards the transition to the phases observed by us.

\section{\label{sec4} Conclusions}
We have presented a stability analysis of the 2D $t-t'$ Hubbard model
on a square lattice for various values of the next-nearest-neighbors
hopping $t'$ and electron concentration. On the basis of the free
energy expression, derived by means of the flow equations method, we
have performed numerical calculation for the various representations
under the point group $C_{4\nu}$ in order to determine the phase
diagram. A surprising large number of phases has been observed. Some
of them have an order parameter with many nodes in ${\bf k}$-space.

For small values of $t'$ and doping the leading instability is the
antiferromagnetic one. At $t'=0$ antiferromagnetism disappears at hole
doping $\delta\sim 0.08$ and temperatures $T\sim 0.1t$. This result
reproduces the N\' eel temperatures in some antiferromagnetic
materials.  The $d_{x^2-y^2}$ wave Pomeranchuk instability dominates
in the region of temperature $T>0.05t$ at small hole doping ($\delta
<0.2$), and for negative $t'\geq -t/3$ the Pomeranchuk instability
dominates in all temperature region at the electron concentrations
around the Van Hove filling. It is the leading instability at large
negative $t'<-t/3$ in the region of electron density above the Van
Hove filling and below some doping; for the value $t'=-5t/12$ this
value of doping equals $\delta\approx 0.2$. The band splitting phase
with staggered $p$-wave symmetry is developed in the region of
electron concentration around half-filling, and is one of the
strongest in that region. This instability dominates when the electron
concentration $n>1.10$ and $t'\neq 0$. The negative
next-nearest-neighbors hopping favors this phase to be dominant
instability at electron doping. The $d_{x^2-y^2}$-wave
superconductivity dominates at low temperatures in certain regions of
electron concentration around half-filling which depend on the value
of $t'\neq 0$. Even large values of $|t'|$ do not destroy the dominant
low-temperature behavior of $d_{x^2-y^2}$-wave superconductivity at
hole doping. It is not destroyed also at sufficiently large hole
doping. However, this dominant behavior is destroyed around the Van
Hove filling where other phases appear to dominate. At small electron
doping $d$-wave superconductivity dominates at low temperatures also,
but as a result of electron-hole asymmetry of the 2D $t-t'$ Hubbard
model the $d$-wave superconductivity appears to dominate within wider
region of hole doping in comparison with the electron doping for
negative values of $t'$.  Triplet and singlet flux phases, which have
degenerate critical temperatures at $t'=0$, occur more easily at the
electron concentrations which are slightly above the Van Hove filling
for $t'\neq 0$. For non-zero values of the next-nearest-neighbors
hopping the critical transition temperatures of these phases are
different, and the triplet one is higher. The triplet analog of flux
phase dominates at low temperatures and $t'=-5t/12$ when the electron
concentration is slightly above the Van Hove filling. The triplet flux
phase is also one of the leading instabilities for various values of
next-nearest-neighbors hopping $t'$ and certain region of electron
concentrations. At $t'<-t/3$ and the Van Hove filling the leading
instabilities are a $g_+$ wave Pomeranchuk instability and $p$-wave
particle-hole instability in triplet channel (phase of magnetic
currents) at temperatures $T<0.15t$, and $s^*$-magnetic phase for
$T>0.15t$. The magnetic currents phase dominates also at very low
temperatures and electron concentrations which are slightly smaller
the Van Hove filling for $t'=-5t/12$, and in the temperature region
$T<0.15t$ below the Van Hove density for $t'=-t/3$.  The
$s^*$-magnetic phase is reduced strongly at low temperatures and shows
a reentrant behavior in certain region of $U/t$. We have found other
weaker instabilities also.  Most instabilities develop at $U>4t$,
which are not small values.  Therefore, flow equation calculations
beyond second order would be desirable.  Nevertheless, as we have
found most commonly discussed types of order, and since some effects
obtained in the intermediate to strong couplings are reproduced
reasonably well by means of the flow equations, we suggest that our
calculations give an estimate of the most important instabilities.
 
\begin{acknowledgments}
We thank E.~Fradkin for his comments and pointing out a Pomeranchuk
instability in nematic Fermi fluid. One of the authors (V.H.) is
indebted to the Institut f\" ur Theoretische Physik for the
hospitality and nice atmosphere during his stay.
\end{acknowledgments}

\newpage


\begin{thebibliography}{99}
\bibitem{iz} Yu.~Izyumov, Usp. Fiz. Nauk {\bf42}, 215 (1999).
\bibitem{sc} D.J.~Scalapino, Physics Reports {\bf250}, 329 (1995).
\bibitem{tk} For a recent review on the cuprate superconductors see 
C.~C.~Tsuei and J.~R.~Kirtley, Rev. Mod. Phys. {\bf72}, 969 (2000).

\bibitem{ha} W.~Hanke {\em et al.}, Adv. Solid State Phys. {\bf38} (1999).

\bibitem{ko} G.~Kotliar, Phys. Rev.~B {\bf37}, 3664 (1988).
\bibitem{am} I.~Affleck, J.~B.~Marston, Phys. Rev.~B {\bf37}, 3774 (1988).

\bibitem{hm} C.~Halboth and W.~Metzner, Phys. Rev. Lett. {\bf85}, 5162 (2000);
Phys. Rev.~B {\bf61}, 7364 (2000).

\bibitem{foh} M.~Fleck, A.~M.~Ole\'{s}, and L.~Hedin, Phys. Rev.~B {\bf56},
3159 (1997).
\bibitem{hsg} R.~Hlubina, S.~Sorella, and F.~Guinea, Phys. Rev. Lett. {\bf78}, 
1343 (1997); Phys. Rev.~B {\bf 59}, 9600 (1999).
\bibitem{ikk} V.~Yu.~Irkhin, A.~A.~Katanin, and M.~I.~Katsnelson, Phys. Rev.~B 
{\bf64}, 165107 (2001).

\bibitem{hs} C.~Honerkamp and M.~Salmhofer, Phys. Rev. Lett. {\bf87},
187004 (2001); Phys. Rev.~B {\bf64}, 184516 (2001).
\bibitem{hsr} C.~Honerkamp, M.~Salmhofer, and T.~M.~Rice, Eur. Phys. J.~B 
{\bf27}, 127 (2002).

\bibitem{gkw} I.~Grote, E.~K\" ording, and F.~Wegner, J. Low Temp. Phys. 
{\bf126}, 1385 (2002).
\bibitem{hgw} V.~Hankevych, I.~Grote, and F.~Wegner, Phys. Rev.~B {\bf66}, 094516 (2002).

\bibitem{hsfr} C.~Honerkamp, M.~Salmhofer, N.~Furukawa, and T.~M.~Rice, Phys. Rev.~B {\bf63}, 035109 (2001).

\bibitem{na} C.~Nayak, Phys. Rev.~B {\bf62}, 4880 (2000).
\bibitem{clmn} S.~Chakravarty, R.~B.~Laughlin, D.~K.~Morr, and C.~Nayak, 
Phys. Rev.~B {\bf63}, 094503 (2001). 

\bibitem{vbh} D.~Vollhardt {\em et al.}, Adv. Solid State Phys. {\bf35}, 383 
(1999).

\bibitem{weg} F.~Wegner, Annalen der Physik (Leipzig) {\bf3}, 77 (1994). 
\bibitem{we} For a recent review on the flow equation method see F.~Wegner, Physics Reports {\bf348}, 77 (2001).

\bibitem{gr} I.~Grote, Ph.D. thesis, University of Heidelberg (2002).

\bibitem{he} C.~Herring, in {\em Magnetism, Vol.~4\/}, ed. by G.~T.~Rao and H.~Suhl (Accademic Press, New York, 1966), p.~1.
\bibitem{fr} J.~Friedel, in {\em The Physics of Metals\/}, ed. by J.~M.~Ziman
(Cambridge University Press, Cambridge, 1969), p.~340.
\bibitem{mot} N.~F.~Mott, {\em Metal-Insulator Transition\/} (Taylor \& 
Francis, London, 1990).

\bibitem{ift} For a review on normal-state properties of high-$T_c$
cuprates see M.~Imada, A.~Fujimori, and Y.Tokura,
Rev. Mod. Phys. {\bf70}, 1039 (1998).

\bibitem{pf} See, for example, P.~Fulde, {\em Electron Correlations in Molecules and Solids\/},
Springer Ser. Solid-State Sci., Vol.~100 (Springer, Berlin, Heidelberg, 1991),
Chap.~14.

\bibitem{we1} F.~Wegner (unpublished).

\bibitem{kk} A.~P.~Kampf and A.~A.~Katanin, cond-mat/0204542.
\bibitem{okf} V.~Oganesyan, S.~A.~Kivelson, E.~Fradkin, Phys. Rev.~B {\bf64}, 195109 (2001). 
\bibitem{kfe} S.~A.~Kivelson, E.~Fradkin, V.~J.~Emery, Nature {\bf393}, 550 (1998). 

\bibitem{mj} S.~Moukouri and M.~Jarrell, Phys. Rev. Lett. {\bf87}, 167010 (2001).
\bibitem{klpt} B.~Kyung, J.~S.~Landry, D.~Poulin and A.-M.~Tremblay, 
cond-mat/0112273.

\bibitem{ho} C.~Honerkamp, Eur. Phys. J.~B {\bf21}, 81 (2001).

\bibitem{hma} T.~C.~Hsu, J.~B.~Marston, I.~Affleck, Phys. Rev.~B {\bf43}, 2866
(1991).

\bibitem{ckn} S.~Chakravarty, H.-Y.~Kee, and C.~Nayak, Int. J. Mod. Phys. {\bf15}, 2901 (2001).
\bibitem{le} P.~A.~Lee, cond-mat/0201052.
\bibitem{va} C.~M.~Varma, Phys. Rev. Lett. {\bf83}, 3538 (1999).


\bibitem{krf} A.~Kaminski {\em et al.}, Nature {\bf416}, 610 (2002).
\bibitem{mo} H.~A.~Mook {\em et al.}, Phys. Rev.~B {\bf64}, 012502 (2001);
{\bf66}, 144513 (2002).
\bibitem{mil} R.~I.~Miller {\em et al.} Phys. Rev. Lett. {\bf88}, 137002 (2002).
\bibitem{tde} For a theoretical discussion of the experiments, see 
Refs.~\cite{hma,ckn} and C.~M.~Varma, Phys. Rev.~B {\bf61}, R3804 (2000).

\bibitem{njk} A.~A.~Narsesyan, G.~I.~Japaridze and I.~G.~Kimeridze, J. Phys.: Condens. Matter {\bf3}, 3353 (1991).
\bibitem{bbd} B.~Binz, D.~Baeriswyl, and B.~Dou\c{c}ot, Eur. Phys. J.~B {\bf25}, 69 (2002).
\end{thebibliography}
\end{document}